\def\year{2020}\relax
\documentclass[letterpaper]{article} %DO NOT CHANGE THIS
\pdfoutput=1
\usepackage{aaai18}  %Required
\usepackage{times}  %Required
\usepackage{helvet}  %Required
\usepackage{courier}  %Required
\usepackage{url}  %Required
\usepackage{graphicx}  %Required
\usepackage{caption}
\usepackage{subcaption}

\usepackage[utf8]{inputenc}   % pacote para acentuação
\usepackage{verbatim}
\usepackage{multirow}
\usepackage{enumitem}

\begin{document}
%\title{A framework to analyze how political polarization affects the behavior of groups related to vaccine stances}
\title{Analysis of the influence of political polarization in the vaccination stance:\\the Brazilian COVID-19 scenario}

\author{Régis Ebeling, Carlos Abel Córdova Sáenz, Jeferson Nobre, Karin Becker\\
\textit{Institute of Informatics} -
\textit{Federal University of Rio Grande do Sul
  (UFRGS)}\\
Porto Alegre, Brazil \\
\{rebeling, cacsaenz, jcnobre, karin.becker\}@inf.ufrgs.br
}
\maketitle

\begin{abstract}

The outbreak of COVID‐19 had a huge global impact, and non-scientific beliefs and political polarization have significantly influenced the population's behavior. In this context, COVID vaccines were made available in an unprecedented time, but a high level of hesitance has been observed that can undermine community immunization. Traditionally, anti-vaccination attitudes are more related to conspiratorial thinking rather than political bias. In Brazil, a country with an exemplar tradition in large-scale vaccination programs, all COVID-related topics have also been discussed under a strong political bias. In this paper, we use a multi-dimensional analysis framework to understand if anti/pro vaccination stances expressed by Brazilians in social media are influenced by political polarization. %if the behavior of Twitter users with distinct stances on COVID-19 vaccination on the Brazilian scenario. The analysis framework incorporates: techniques to automatically infer from users their demographics and political orientation, topic modeling to discover the concerns expressed by each group, network analysis to characterize their behavior as a social network group, analysis of linguistic characteristics to identify psychological aspects, and the characterization of information sources and external influence to evaluate the ``echo chamber effect''. We identified one pro-vaccination group, two anti-vaccination groups, and a neutral group. 
The analysis framework incorporates: techniques to automatically infer from users their 
%demographics and 
political orientation, topic modeling to discover their concerns, network analysis to characterize their social behavior, %analysis of linguistic characteristics to identify psychological aspects, 
and the characterization of information sources and external influence. % to evaluate the ``echo chamber effect''. 
Our main findings confirm that anti/pro-stances are biased by political polarization, right and left, respectively. While a significant proportion of pro-vaxxers display haste for an immunization program and criticize the government's actions, the anti-vaxxers distrust a vaccine developed in a record time. Anti-vaccination stance is also related to prejudice against China (anti-sinovaxxers), revealing conspiratorial theories related to communism. All groups display an ``echo chamber" behavior, revealing they are not open to distinct views. 

\end{abstract}

\section{Introduction}

% 1

Since the outbreak in China in late 2019, the novel coronavirus disease (COVID‐19) has quickly spread around the world. Due to the pandemic's urgency, efforts were undertaken to develop COVID-19 vaccines, approve and make them available in the shortest possible time frame. These efforts had to follow the same legal requirements for pharmaceutical quality, safety, and efficacy as other medicines. COVID-19 immunization programs have started in Europe by the end of 2020. 

% 2

Despite being recognized as one of the most successful public health measures, a growing number of people perceive the vaccination as unsafe and unnecessary \cite{Hornsey2018}. Anti-vaccination movements have been implicated in lowered vaccine acceptance rates and the increase in vaccine-preventable disease outbreaks. Studies indicate that anti-vaccination attitudes are more related to conspiratorial thinking rather than political bias or religious beliefs~\cite{Hornsey2018,Bryden2019,Cossard2020}. In this context, SARS-CoV-2 vaccines have been the target of all sorts of fake news and misinformation \cite{Matamoros2020}. Vaccine hesitancy and misinformation present substantial obstacles to achieving coverage and community immunization ~\cite{Burki2020}. Brazil has a successful history of immunization programs, and its National Immunization Program (NIP) is a world-class reference on the large-scale eradication of many diseases \cite{Domingues2020}.

Political polarization has influenced the population's behavior towards COVID-19, including vaccination. In the United States of America (US), studies \cite{Makridis2020,Bruine2020} reveal that partisan affiliation is often the strongest single predictor of behavior towards COVID-19. Such political polarization was also observed in Brazil, a country politically divided since the 2018 Presidential election, when the right-wing candidate Jair Bolsonaro was the winner. Political polarization has been a major obstacle in the planning and implantation of a national COVID-19 immunization program, despite the large experience with the NPI and the existence of national research centers able to produce vaccines  (e.g., Fiocruz, Butantan). In addition, substantial efforts were not undertaken throughout 2020 to secure contracts with the international pharmaceutical industries to buy vaccines.

At the same time, Brazilian governors and mayors, who have to deal with day-by-day COVID-19 demands directly, are concerned about the collapse of the health system and exert pressure for large-scale vaccination. Mandatory vaccination is defended by many of them to reach community immunization, so as to preserve lives and recover the economy. Joao Doria, governor of São Paulo, has devised a state immunization program, securing funds to develop the vaccine Coronavac in a joint effort between Instituto Butantan and the Chinese pharmaceutical company Sinovac. Since Doria is a prospective candidate for the 2022 Presidential election, Bolsonaro has repeatedly undermined Doria's efforts regarding Coronavac, using derogatory and xenophobic terms to refer to it, as well as making direct offenses to China (which resulted in the delay of Chinese pharmaceutical inputs to produce Coronavac). The ``Phony War" between Bolsonaro and Doria has been fought mostly on social media. The Brazilian vaccination program has started in late January 2021 with Coronavac.

Significant research efforts have addressed COVID-related discourse in social media.  Regarding the influence of political polarization, \cite{Jiang2020} examines geographic differences in online discourses, \cite{sha2020} analyzes narratives according to governmental decision making, \cite{Rao2020} investigates its relationship with anti-science behavior, and \cite{ebling2020} characterizes the influence of political polarization in groups pro/against social distance. Misinformation about COVID is addressed in works such as \cite{Cinelli2020,Furini2020,Burki2020}, and the ideological influence is stressed in \cite{Havey2020}. Regarding COVID-19 vaccination on social media, a few pre-print works propose topic modeling techniques to provide insights on anti-vaccination views  \cite{lyu2020,curiel2020}. To the best of our knowledge, existing works were not successful in relating anti-vaccination behavior to political ideology (e.g., \cite{Hornsey2018,Czarnek2020}).

% 6

In this paper, we use a multi-dimensional analysis framework to understand Twitter users' behavior with distinct stances on COVID-19 vaccination and their relationship with the political polarization in the Brazilian scenario. The analysis framework incorporates: a) topic modeling to discover the specific concerns expressed by each group and their possible political bias; b) an index to infer the political orientation of users based on the politicians followed, c) network analysis and community detection to characterize the effect of political polarization in the behavior as a social network group; and d) the characterization of information sources and external influence to evaluate the ``echo chamber effect". We analyze and compare a pro-vaccination group, two anti-vaccination groups, and a neutral group. We divided into two anti-vaccination groups because we identified a significant number of users expressing derogatory ideas specifically related to Coronavac. In this work, these groups are referred to as \textit{pro-vaxxers}, \textit{anti-vaxxers} and \textit{anti-sinovaxxers}. The latter is a subgroup of anti-vaxxers, in which the anti-stance is not limited to the vaccination itself: it also encompasses the partnership with the Chinese Sinovac and Bolsonaro's political rival Joao Doria.

In regards to the Brazilian scenario of COVID-19 vaccination, we aim to answer the following research questions:   

\begin{itemize}
[leftmargin=0.3cm,topsep=0pt,itemsep=-1ex,partopsep=1ex,parsep=1ex]
\vspace{-3pt}
\item Q1: Is there a difference in the topics discussed by each group?
\item Q2: Are these groups politically polarized? 
\item Q3: Does the polarization affect the social network structure of these groups?
%\item Q3: Do these groups have different psychological aspects?
\item Q4: Is there a difference in the information sources used by each group?
\end{itemize}

Our main findings confirm that anti/pro stances in the Brazilian COVID vaccination scenario are biased by political polarization, right and left, respectively. A significant proportion of pro-vaxxers display haste for an immunization program and criticize the President and the government's actions. The anti-vaxxers distrust a vaccine developed in a record time and defend immunization as an individual choice. Anti-vaccination stance is also related to prejudice against China (anti-sinovaxxers), revealing conspiracy theories related to communism and support to the rivalry between prospective candidates for the 2022 Presidential election. As a group, the social networks of anti-sinovaxxers and anti-vaxxers are more connected and act as spokespersons of right-wing politicians, while mainstream media and science activists %socially
influence the pro-vaxxers. All groups display an ``echo chamber" behavior, revealing they are not open to distinct views but rather seek to reinforce their convictions.

The remainder of this paper is structured as follows. Section 2\ref{sec:trabalhosRelacionado} discusses related work. Section 3\ref{sec:materiaisEMetodos} discusses the data used and the techniques deployed for the proposed analysis. Section 4\ref{sec:analises} presents the analyses developed to answer the research questions. Section 5\ref{sec:threats} discusses the validity threats to our study. Section 6\ref{sec:conclusoes} draws conclusions and future work.
\section{Related Work}

\label{sec:trabalhosRelacionado}
Twitter has been used to study different social phenomena, such as gender equality ~\cite{Elsherief2017},
racial equity \cite{DeChoudhury2016} or political polarization \cite{Garimella2017}. %Information extracted automatically from the users' profiles allows deepening the understanding of these phenomena according to demographics, such as the gender/age inferred from profile pictures~\cite{Elsherief2017}, or the political orientation based on the politicians followed~\cite{Venkata2017}.

Different works have extracted information from user's profiles, their social structure, or information spread in the network to make predictions of political partisanship. Works such as \cite{Barbera2015,Garimella2017} derive a user's political polarization by analyzing the users they follow in the social network. Political polarization can also be inferred by analyzing clusters of retweets/mentions of tweets from users \cite{Conover2011} or by models analyzing a set of features extracted from the tweets \cite{PreotiucPietro2017}. 

COVID-19 in social media is a very active research area, and we can find more than 150 pre-print works in arXiv. Many works investigate online conversations in terms of topics, information diffusion, and topics change overtime~\cite{ordun2020exploratory,garcia2020}. Regarding political polarization, \cite{Jiang2020} examines geographic differences in online COVID-19 discourse, relating the polarization to each US state's political dominance. \cite{sha2020} presents a longitudinal study relating Twitter narratives to Governors and Presidential actions. \cite{Rao2020} examines the ideological alignment of users along moderacy, political and science dimensions, concluding that moderacy is the key influence on science's behavior. \cite{Havey2020} relates misinformation topics about the COVID-19 with political ideology, concluding that liberals are more prone to believe/spread misinformation.
% Jeferson melhor em uma frase para ser parecido com as outras citações
\cite{ebling2020} analyzed multiple dimensions of the political influence in pro/anti social distance stances. They concluded that these groups display similar psychological aspects, but liberals tend to form a more cohesive, socially connected group that are not interested nor open to other views.

The far-from-universal willingness to accept a COVID-19 vaccine represents substantial obstacles to achieving coverage and community immunity \cite{Lazarus2020}. The prevalent reasons for anti-vaccination attitudes are conspiratorial thinking, followed by a low tolerance for impingement on their freedoms~\cite{Hornsey2018}. Although studies show that political partisanship is the strongest single predictor of behavior regarding COVID~\cite{Makridis2020,Bruine2020}, to the best of our knowledge, ideology has not found to be influential on the anti-vaccination stance. Studies reveal that while anti-vaxxers will not change their view under any argument, the pro-vaccine groups have been reactive and reticent due to legitimate questions~\cite{Burki2020}. According to \cite{Cossard2020}, vaccination skeptics and advocates reside in their own distinct ``echo chamber", but their network information structure is distinct. 
Insights in the doubts and misinformation spreading have been addressed by works such as \cite{lyu2020,curiel2020,Matamoros2020,Cinelli2020,Kaliyar2021}. 

Our study differs from related work by examining the political influence on pro/anti-vaccination stances in the Brazilian COVID scenario. We investigate Twitter users according to an analysis framework that combines topic modeling for summarizing concerns, political polarization measuring, analysis of their social network structure, and information sources.

%\section{Analysis Framework on the Influence of Polarization in Groups Behavior}
\section{Analysis Framework}
\label{sec:materiaisEMetodos}

\begin{figure}[!t]
	\centering
    \includegraphics[width=0.72\columnwidth]{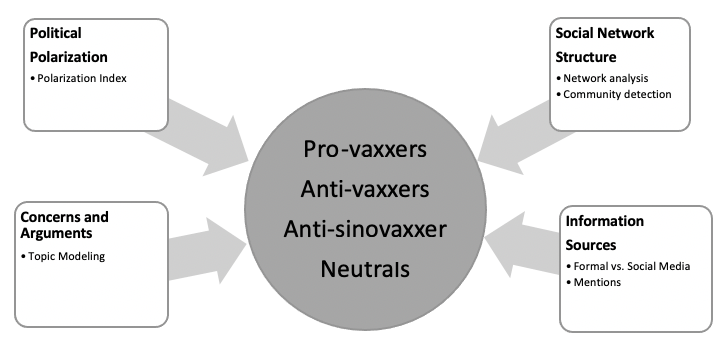}
%	\vspace{-10pt}
	\caption{Analysis Framework}
	\label{fig:analysisFramework}
\end{figure}

This paper investigates the political polarization influence on the vaccination stances expressed by Brazilians on Twitter. We identified distinct groups with anti/pro stances and deployed a multi-dimensional framework that encompasses %demographics, 
identification of concerns, polarization level, social network properties, and information sources. Figure \ref{fig:analysisFramework} displays the techniques associated with each dimension, detailed in the remainder of this section. 

\begin{table}[!t]
\centering
\caption{Hashtags and collection numbers per group}
\label{tab:coletaDados}
\scalebox{0.9}{
\begin{tabular}{|l|l|l|l|l|l}
\textbf{Group} & \textbf{Nº Tweets} & \textbf{Nº Users}\\
 \hline
Pro-vaxxers & 160.867 & 100.847\\
Anti-vaxxers & 32.876 & 15.647\\
Anti-sinovaxxers & 17.810 & 7.067\\
Neutrals & 19.558 & 18.396\\
\end{tabular}
}
\end{table}

\begin{figure*}[!ht]
	\centering
	\includegraphics[width=1.8\columnwidth]{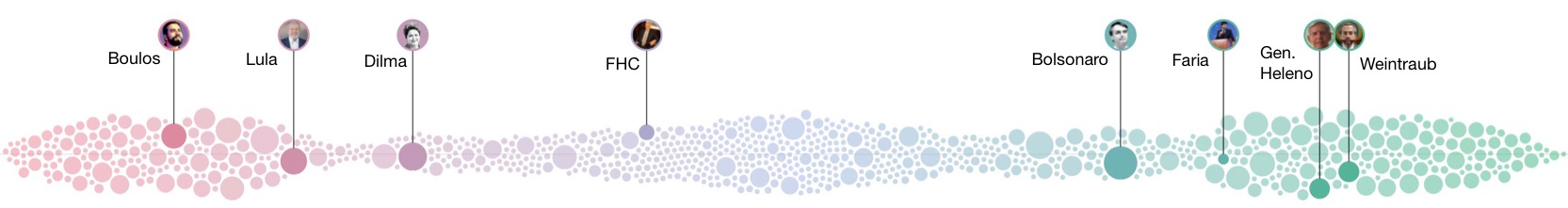}
	\caption{iGPS with Brazilian Presidents and Political Influencers}
	\label{fig:igsp}
\end{figure*}

\subsection{Data Crawling and Groups}
\label{sec:dados}

To compose the groups with distinct vaccination stances, we crawled tweets containing the terms ``vaccine" or ``vaccination" in Portuguese, from January 1\textsuperscript{st} 2020 to April 1\textsuperscript{st} 2021. This period covers the pandemic since its beginning, all phases of vaccine development and approval, as well as the first trimester of vaccination in 2021. We collected a total of 236.992 tweets. Then, we extracted all hashtags, ordered by frequency, and grouped them according to the represented stance. We identified four distinct groups according to the meaning of the hashtags:

\begin{itemize}
[leftmargin=0.3cm,topsep=0pt,itemsep=-1ex,partopsep=1ex,parsep=1ex]
    \item \textit{Pro-vaxxers}: this group represents people who express a pro-vaccination stance, by using hashtags that express endorsement for immunization programs (e.g., VaccinesForEverybody), immunization intention (e.g., IWillReceiveVaccine) or immunization urgency (e.g., VaccineNow). The portuguese hashtags are \#EuVouTomarVacina, \#VacinaBrasil, \#VacinaÉAmorAoPróximo, \#VacinaJá, \#VacinaNoBrasil, \#VacinaParaTodos, \#VacinasPelaVida, \#VacinaUrgenteParaTodos and \#VemVacina; 
     \item \textit{Anti-vaxxers}: this group represents people who took a stance against COVID-19 vaccination, using hashtags expressing they do not want to get vaccinated (e.g., NoVaccination) or are againts mandatory vaccination (e.g.. NoMandatoryVaccine). The hashtags that represent this group are \#EuNãoVouTomarVacina, \#NãoVouTomarVacina, \#VacinaNão and \#VacinaObrigatóriaNão;  
     \item \textit{Anti-sinovaxxers}: among the anti-vaxxers, we found opinions specifically against Coronavac. The hashtags involve derogatory expressions related to China (e.g. ``vacchina") and are consistent with the contempt display by Bolsonaro's in regards to Doria's effort for an immunization program. Thus, we decided to create a specific group to analyze whether the anti-stance represented in this group differs from the general anti-vaccine stance of the Anti-vaxxers. The hashtags are \#VachinaNão, \#VachinaNãoPresidente, \#VachinaObrigatóriaNão and \#VacinaChinesaNão;      
    \item \textit{Neutrals}: to represent this group, we selected tweets with ``vaccine" or ``vaccination", no hashtags, and no mentions. The volume collected at random to represent this group was the average number of tweets of the other groups.
\end{itemize}

 Table \ref{tab:coletaDados} displays the volume of collected and preprocessed tweets and the respective number of users. We crawled the data using the library snscrape API\footnote{https://pypi.org/project/snscrape/}. We excluded bots identified using the API Botometer\footnote{https://rapidapi.com/OSoMe/api/botometer-pro}, as well as suspended users. We discarded tweets with less than three terms. 

We applied classic pre-processing actions for topic modeling, such as case normalization and removal of stop words, punctuation, special characters, hashtags, URLs, and user mentions.

\subsection{Topic Modeling}
To investigate if there is a difference in the topics discussed by each group (Q1), we combined two topic modeling techniques: Latent Dirichlet Allocation (LDA)~\cite{blei2003latent} and BERTopic~\cite{bertopic2020}. We regard these techniques as complementary since LDA provides a coarse-grained clustering based on probabilities of words co-occurrence in a documents corpus, and BERTopic helps to identify frequent similar arguments based on tweet similarity. Finding topics at two granularity levels enables us to leverage the strengths of each technique while reducing their drawbacks. LDA divides the groups in terms of general concerns, and this reduced search space enables us to identify representative arguments used to support the stances using BERTopic.

The LDA input is the corpus and a parameter $k$ (number of topics to discover). The output is a set of $k$ topics, consisting of terms and probabilities.  To evaluate the results, we used the metric CV~\cite{roder2015exploring}, which measures the coherence of topics based on multiple dimensions. We applied LDA on the set of pre-processed tweets for each group. We varied $k$ from 1 to 30 to find the best $k$ for each case, selecting the ones with the best CV values. Then, we manually inspected the results using the most representative terms for each topic and a sample of associated tweets. The $k$ chosen for each group represents the smallest set of topics found with a coherent set of terms and the least redundancy topics. Using Gensim\footnote{https://pypi.org/project/gensim/} ($alpha=0.5$, $beta=auto$), we chose $k=3$ for Anti-sinovaxxers, $k=4$ for Anti-vaxxers and $k=5$ for Neutrals and Pro-vaxxers. 

BERTopic is a framework encompassing algorithms to automatically seek dense topics in a collection of documents, assuming that semantically similar documents form topics within the input collection. It requires the input of a corpus and a pre-trained language representation model (e.g., BERT). After dimensionality reduction, it finds dense areas of similar documents in the vector space using HDBScan, a density-based clustering algorithm. Unlike LDA, BERTopic does not require as input the number of clusters. However, it has as a drawback the huge number of resulting clusters, which jeopardizes the interpretation of their semantics.

We propose to find the most representative arguments for each one of the topics found using LDA. We input to BERTopic the set of documents associated to a given LDA topic and find clusters of similar tweets. To identify the most representative tweet of each cluster, we search for the tweet with the highest average similarity with regard to all tweets of the same cluster. We analyzed the three largest clusters for each topic of each group. We used BERTopic library\footnote{https://pypi.org/project/bertopic/0.3.4/} and the `distiluse-base-multilingual-cased'\footnote{https://huggingface.co/distilbert-base-multilingual-cased}, a BERT trained model that supports the use of 50 different languages.

\subsection{Political Polarization Index}
\label{sec:polarizationIndex}

To investigate if users are politically polarized (Q2), we propose an index to measure the political polarization of the users according to the right/left politicians they follow. 
For each user, we collect the list of users followed (\textit{followings}). Then, we calculate the ratio between the number of followed right-oriented politicians and the total number of politicians followed (right or left-oriented). We adopted an offset of 1 for each side to adjust the calculation when a user does not follow right or left politicians. Thus, the value 50\% indicates politically neutral users, i.e., either do not follow politicians or follow them in equal amounts. The higher (or lower) the metric value, the more oriented to the right (left) the user is.

We adopted the Ideological GPS\footnote{http://temas.folha.uol.com.br/gps-ideologico/} (iGPS) to select the right/left-oriented politicians. The Ideological GPS is a tool that adapts the well-known statistical model proposed by Barberá et al. \cite{Barbera2015}, which calculates political polarization using the followers' structure. The iGPS analyzed political influencers, such as artists, politicians, journalists, etc. Figure \ref{fig:igsp} shows the position of Brazilian presidents in the iGPS, such as Lula, Dilma Roussef, Fernando Henrique Cardoso (FHC), and Jair Bolsonaro. The figure also shows influential politicians who appeared in our study. Note that the left-oriented politicians, represented in red, are separated from the right-oriented politicians (represented in green) by a neutral zone, depicting moderate (neutral) influencers. For example, FHC is regarded as a moderate who has echo on both sides of the spectrum. %, as illustrated in the distribution of Figure \ref{fig:igsp}. 
We selected the 165 most left-oriented politicians and the 165 most right-oriented ones, which corresponds to the left/right-most politicians outside the borders of the neutral zone. 

\subsection{Polarization Influence in the Social Network Structure}

We analyzed the influence of polarization in the social network structure (Q3) based on the structural relationships defined by the following list. For each group, we constructed a directed graph where the nodes correspond to users from the group or followed by them, and directed edges connect  nodes according to the followings list. First, we calculated global measures for each network to characterize their complexity, such as degree, average shortest path, diameter, and clustering coefficient. Next, we detected communities (i.e., subgroups) within each network using the Gephi software\footnote{https://gephi.org/}. For each community, we calculated the same metrics, as well as closeness centrality (i.e., how close a node is to others) and the betweenness centrality (i.e., quantifying the number of times a node acts as a bridge for other pairs of nodes). 

Finally, we assessed the political polarization of the identified communities using the same list of politicians adopted for the polarization index. We identified two polarized communities for each group, which essentially included politicians of a single political orientation (i.e., only right or left), with a few exceptions. All the other communities did not include politicians (or virtually none). For these polarized communities, we inspected the strengths of the connections between different users using subgraphs that connected them. We examined the connections within graphs containing politicians only (left and right) and connections between subgraphs containing regular users connected to politicians (left and right). In this way, we can evaluate the political influence in the spread of information in each community.

Since the software used could not handle the size of the two biggest graphs, corresponding to the Anti-vaxxers and Pro-vaxxers, we adopted the following sampling method. For the Anti-vaxxers, we randomly divided the group into three folds of users and constructed the respective graphs three times using pairs of folds. For each graph, we calculated all metrics and identified the communities. Then, we compared the results of the three graphs. The graphs varied in terms of the number of nodes (1.4M-1.55M) and edges (5M-5.5M) but were very similar in terms of communities found (average 44, STD=1.73). In two samples, we identified polarized communities with similar properties, with identical centrality nodes for the polarized communities. For the Pro-vaxxers, we applied the same sampling method, but due to the amount of data, we divided the data into ten folds, repeating the process five times. The graphs varied in terms of the number of nodes (1.8M-2.1M) and edges (5.4M-5.98M) and were somewhat similar in terms of communities (average 74.6, STD=8). However, we found similar polarized communities in 4 graphs, both in terms of topological metrics and centrality nodes.  
Thus we selected the samples that yielded the graphs with the most similar topological metrics, including a similar number of left/right politicians in the polarized communities and the same centrality nodes. 

\subsection{Information Sources}
To analyze if there exist differences in the source of information exchanged by the groups (Q4), we used the URLs and mentions in the tweets. Links represent the sources of information used to ground an opinion or spread information to other group members. A mention in a tweet represents that another person's opinion is relevant to the discussion. This importance may translate on the spread, endorsement, discussion, or opposition of/to other people's points of view.  

We divided the URLs into three subcategories: verified news portals, social network addresses, and others. We classified the news portals using the indexed list available in Kadaza\footnote{https://www.kadaza.com.br/noticias}, which contains 21 recognized news outlets. We also inspected the list of the top-30 most frequent URLs referenced in the tweets, looking for neutral news outlets. Although some URLs did correspond to widely well-known news websites, all of them were associated with a clear far-right/left point of view. For neutrality, we did not classify them as news portals. Addresses related to social networks were classified as Instagram, Facebook, and Youtube. We did not found significant references to fact-checking sites, and thus we disregarded this category. The analysis did not consider all other websites.

Mentions were classified according to group membership: Anti-vaxxers, Anti-sinovaxxers, and Pro-vaxxers. We also identified users as left/right politicians from our ideological GPS list. To avoid bias, we identified the intersection of users in distinct groups, finding 851 common users between the Anti-sinovaxxers and Pro-vaxxers and 1768 common users between the Anti-sinovaxxers and Anti-vaxxers. The intersection of the three groups has 340 users. In the analysis of mentions, each user's behavior is considered only once (e.g., in the analysis of Anti-vaxxers, a user belonging to more than one group is regarded only as Anti-vaxxer).  

\section{Results} 
\label{sec:analises}

\subsection{Q1: Is there a difference in each group's topics discussed?}
\label{sec:assuntos}

\begin{table}[!t]
\centering
\vspace{-12pt}
\caption{Topics per group}
\vspace{-5pt}
\label{tab:atributosTopicos}
\scalebox{0.7}{
\begin{tabular}{llllll}
\hline
\textbf{Top.}  & \textbf{Tweets} & \textbf{Users} & \textbf{Dens.} & \textbf{BT Cl.} & \textbf{Anti-vaxxers: Words}                                   \\ \hline
0  & 25,9\% & \textbf{40,6\%} & 1,34 & 164 & body, rules, respect, want
        \\ 
1  & 21,8\% &  28,9\% & 1,58 & 147 & vaccine, lab rat, liberty, Bolsonaro                                    \\ 
2  & 20,2\% &  25,4\% & 1,66 & 133 & get, point, final, dictators \\                            
3  & \textbf{32,1\%} &  36,6\% & \textbf{1,84} & 173 & Brazil, vaccine, people, Doria \\
\hline
\textbf{Top.}  & \textbf{Tweets} & \textbf{Users} & \textbf{Dens.} & \textbf{BT Cl.} & \textbf{Anti-sinovaxxers: Words }                                  \\ \hline
0  &  30,9\% & 40,1\% & 1,94 & 113 & \begin{tabular}[c]{@{}l@{}l@{}l@{}}doria, dictator, china, want\end{tabular} \\
1  &  29,9\% &  47,7\% & 1,58 & 104 & china, get, this, brazil                                         \\
2  & \textbf{39,1\%} & \textbf{48,8\%} & \textbf{2,01} & 155 & vaccine, chinese, president, against \\
\hline
\textbf{Top.}  & \textbf{Tweets} & \textbf{Users} & \textbf{Dens.} & \textbf{BT Cl.} & \textbf{Pro-vaxxers: Words}                                   \\ \hline
0  & 18,2\% & \textbf{18,8\%} & 1,54 & 510 & brazil, approved, people, vaccine       
        \\ 
1  & 16,8\% & 16,9\% & 1,57 & 467 & ready, get, waiting, alligator                                    \\ 
2  & 18,8\% & 15,6\% & 1,92 & 485 & science, bolsonaro, out, vaccination \\
3  & 19,9\% & 16\% & 1,98 & 508 & life, happy, vaccinated, let's go \\
4  & \textbf{26,3\%} & 18,4\% & \textbf{2,28} & 665 & coronavac, anvisa, today, first \\
\hline
\textbf{Top.}  & \textbf{Tweets} & \textbf{Users} & \textbf{Dens.} & \textbf{BT Cl.} & \textbf{Neutrals: Words}                                   \\ \hline
0  & \textbf{24,2\%} & \textbf{25\%} & \textbf{1,03} & 534 & covid, where, get, want
        \\ 
1  & 22,5\% &  23,5\% & 1,01 & 485 & when, take, arm, to take                                    \\ 
2  & 19,8\% &  20,5\% & 1,02 & 454 & against, chinese, can, tomorrow \\                            
3  & 18,5\% &  19,4\% & 1,01 & 467 & want, TGIF, walk, soon \\
4  & 14,9\% &  15,6\% & 1,01 & 412 & cure, god, invest, cheap \\
\end{tabular}
}
\end{table}

\begin{table}[!t]
\centering
\caption{Anti-vaxxers: three largest clusters}
\label{tab:clusterArgumentosAntivax}
\vspace{-5pt}
\scalebox{0.7}{
\begin{tabular}{l|l|l|l}
\textbf{Top.} & \textbf{Clust.} &  \textbf{\#Tw.} & \textbf{Representative Argument}\\
 \hline
& A00 & 155 & ``Brazil is going to show that we are not \\
& & & STF's puppets"\\
0 & A01 & 131 & ``My body, my choice"\\
& A02 & 141 & ``Mandatory vaccination my @\#!"\\ 
\hline
 
& A10 & 138 & ``Bolsonaro always"\\
1 & A11 & 114 & ``All of Brazil in Brasilia" \\
& A12 & 91 & ``Noone is obliging to to take this s\#!@"\\ 
\hline

& A20 & 250 & ``Great, that's what natural selection is for" \\
2 & A21 & 215 & ``Wake up Brazil" \\
& A22 & 118 & ``I raise a prayer to God amem" \\ 
\hline

& A30 & 636 & ``NO to mandatory vaccination, we will not be \\ 
& & & China's lab rats" \\
3 & A31 & 281 & ``Let's do as the people from Buzios and raid Brasilia" \\
& A32 & 253 & ``Let them be lab rats" \\ 
\hline
\end{tabular}
}
\end{table}

\begin{table}[!t]
\centering
\caption{Anti-sinovaxxers: three largest clusters}
\label{tab:clusterArgumentosAntisino}
\vspace{-5pt}
\scalebox{0.7}{
\begin{tabular}{l|l|l|l}
\textbf{Top.} & \textbf{Clust.} &  \textbf{\#Tw.} & \textbf{Representative Argument}\\
 \hline
& S00 & 409 & ``Doria dictatorship. Your corruption vaccine \\
& & &  won't work on me"\\
0 & S01 & 251 & ``We'll show that we won't get vaccinated with  \\
& & & chinese vaccines, we have the power to fight it"\\
& S02 & 147 & ``Bolsonaro: best president in the history of Brazil"\\ 
\hline
 
& S10 & 518 & ``I will never take Doria's vaccine"\\
1 & S11 & 212 & ``No, China never ever" \\
& S12 & 88 & ``Man, you have a lot of courage to deny a vaccine \\
& & &  because it's from China. I'm already waiting  \\
& & & in line  even if it comes from Chernobyl"\\ 
\hline

& S20 & 228 & ``Say no to the Chinese vaccine Doria wants to force on \\
& & & São Paulo's people" \\
2 & S21 & 154 & ``Brazilians are not lab rats" \\
& S22 & 100 & ``You guys can have these vaccines made in such a rush" \\ \hline
\end{tabular}
}
\end{table}

\begin{table}[!t]
\centering
\caption{Pro-vaxxers: three largest clusters}
\label{tab:clusterArgumentosProvax}
\vspace{-5pt}
\scalebox{0.7}{
\begin{tabular}{l|l|l|l}
\textbf{Top.} & \textbf{Clust.} &  \textbf{\#Tw.} & \textbf{Representative Argument}\\
 \hline
& P00 & 483 & ``Cheers to SUS, cheers to scientists, cheers to\\
& & & Butantan, cheers to Fiocruz, cheers to the vaccine"\\
0 & P01 & 425 & ``Butantan is serving Brazil, working to save lives"\\
& P02 & 326 & ``I waited so long for this moment"\\ 
\hline

& P10 & 757 & ``Brazil, here comes the vaccine"\\
1 & P11 & 538 & ``I am so happy, so much joy, vaccines have come" \\
& P12 & 339 & ``I am ready to get any vaccine"\\ 
\hline

& P20 & 467 & ``I will get vaccinated for myself and for \\
& & & all Brazilians" \\
2 & P21 & 374 & ``Until when will we pay for bolsonarist \\ 
& & & incompetence? Vaccines for all" \\
& P22 & 319 & ``It's revolting to have someone so unprepared \\
& & & as president" \\ 
\hline

& P30 & 534 & ``The vaccines vaccines in Brazilian soil" \\
3 & P31 & 405 & ``Has anyone seen Bolsonaro? He vanished after\\
& & & they announced vaccine approval" \\
& P32 & 321 & ``You garbage, you are a disgrace to this country,\\ 
& & & you support genocide" \\ 
\hline

& P40 & 969 & ``My greatest wish today: urgent vaccines for all\\ 
& & & the population so schools can open again" \\
4 & P41 & 618 & ``I want to get vaccinated, I want to be free from the \\
& & & threat of 'coronga' so I can fight against the threat \\
& & & of Bolsonaro" \\
& P42 & 300 & ``And so the plan of parliamentary immunity and\\
& & & genocide goes on in Brazil" \\ 
\hline
\end{tabular}
}
\end{table}

\begin{table}[!t]
\centering
\caption{Neutrals: largest cluster}
\label{tab:clusterArgumentosNeutrals}
\vspace{-5pt}
\scalebox{0.8}{
\begin{tabular}{l|l|l|l}
\textbf{Top.} & \textbf{Clust.} &  \textbf{\#Tw.} & \textbf{Representative Argument}\\
 \hline
0 & N00 & 356 & ``Where is the vaccine?"\\
\hline

1 & N10 & 223 & ``Brazil has 2.4 million people vaccinated. \\
& & & USA vaccine 2 million PER DAY."\\
\hline

2 & N20 & 339 & ``Most at-risk group, without mask and\\ 
& & & crowding, can get out without vaccine" \\
\hline

3 & N30 & 299 & ``Am I living or just waiting for the\\ & & & vaccine? TGIF" \\
\hline

4 & N40 & 312 & ``When the vaccine is available to you,\\ 
& & & VACCINE!" \\
\hline
\end{tabular}
}
\end{table}

We analyzed the topics discussed by each group and the representative tweets of their arguments to express their point of view. We elaborate the following hypotheses: a) the concerns expressed by Anti-vaxxers, Anti-sinovaxxers, and Pro-vaxxers groups go beyond expressing stance about vaccination, embedding a significant political bias; b) Anti-sinovaxxers and Anti-vaxxers represent different anti-vaccination standpoints. 

\begin{figure}[!t]
	\centering
	\includegraphics[width=0.7\columnwidth]{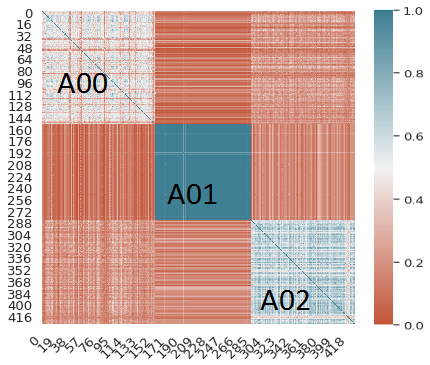}
	\caption{Similarity Matrix - Topic 0 (Anti-vaxxers)}
	\vspace{-10pt}
	\label{fig:matrizSimilaridadeA0}
\end{figure}

\begin{figure}[!t]
	\centering
	\includegraphics[width=0.7\columnwidth]{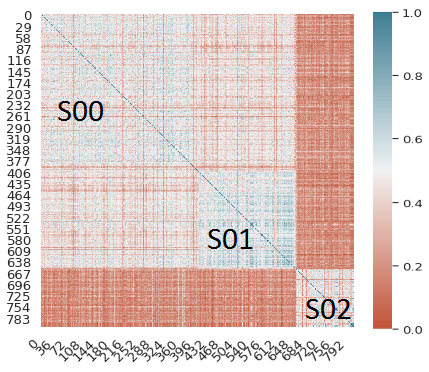}
	\caption{Similarity Matrix - Topic 0 (Anti-sinovaxxers)}
	\vspace{-10pt}
	\label{fig:matrizSimilaridadeS0}
\end{figure}

\begin{figure}[!t]
	\centering
	\includegraphics[width=0.7\columnwidth]{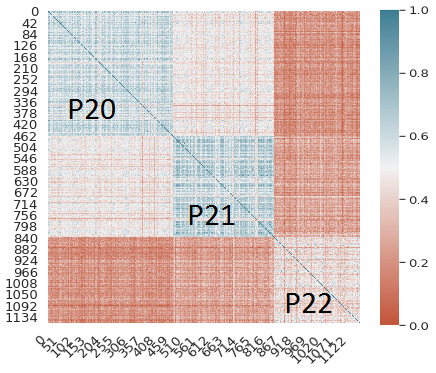}
	\caption{Similarity Matrix - Topic 0 (Pro-vaxxers)}
	\vspace{-10pt}
	\label{fig:matrizSimilaridadeP2}
\end{figure}

Table \ref{tab:atributosTopicos} shows the LDA topics identified for each group, described by the percentage of tweets and users, density (tweets/user), number of BERTopic clusters found, and the four most representative words (highest weight). We interpreted each topic's meaning by inspecting the top-50 influential words and the manual inspection of a sample of related tweets. Then, we further refined this interpretation using the central arguments of BERTopic clusters. Tables \ref{tab:clusterArgumentosAntivax}, \ref{tab:clusterArgumentosAntisino} and \ref{tab:clusterArgumentosProvax} present the clusters of semantically similar arguments for the Anti-vaxxers, Anti-sinovaxxers and Pro-vaxxers. For the three biggest clusters of each topic, the tables summarize the number of tweets and provide a representative tweet (i.e., greatest similarity to the other tweets of the same cluster). For each group, we provide as example in figures \ref{fig:matrizSimilaridadeA0}, \ref{fig:matrizSimilaridadeS0} and \ref{fig:matrizSimilaridadeP2} the similarity matrix of the clusters identified for one topic (Anti-vaxxers, Anti-sinovaxxers, and Pro-vaxxers, respectively). Table \ref{tab:clusterArgumentosNeutrals} presents representative arguments for the Neutrals group for the largest cluster. In the remaining of this section, we describe the results for each group and discuss our results. 

\noindent \textbf{Anti-vaxxers. } the topics in Table \ref{tab:atributosTopicos} reveal that users in this group are concerned about their freedom regarding vaccines that they do not trust, as summarized by the representative arguments in Table \ref{tab:clusterArgumentosAntivax}. To understand the central arguments of this group, it is necessary to highlight the conflict between the President and the governors. While the former repeatedly spread in social media contempt about vaccination, many governors (including Doria) defended that mandatory vaccination is required in Brazil for sanitary reasons, raising an inflamed debate about constitutionality. By December 2020, the Federal Supreme Court (STF) ruled that mandatory vaccination is constitutional. \textit{Topic 0} addresses issues related to mandatory vaccination. The main arguments are the complaints regarding the STF ruling about mandatory vaccination (A00), the mantra ``my body, my choice" (A01), and swearing about mandatory vaccination (A2). The similarity matrix comparing these clusters is displayed in Figure \ref{fig:matrizSimilaridadeA0}. \textit{Topic 1} questions the safety of vaccines approved in such a short time. The main arguments are praises to the President (A10), organization of protests in Brasilia against the STF (A11), and outrage against mandatory vaccination (A12). \textit{Topic 2} expresses discontent about what is perceived as authoritarianism (i.e., the STF ruling and the pressure for mandatory vaccination by governors). Interestingly, the biggest cluster (A20) endorses vaccination by reminding those who are not willing to get vaccinated is subject to Darwin's natural selection, which actually reveals a heated discussion between anti/pro-vaxxers.
The other representative arguments raise awareness about the STF/governor's dictatorship (A21) and praise for God's protection to the President (A22). Finally, \textit{Topic 3} addresses the rivalry with Doria and the right to freedom guaranteed by the Constitution. The most representative argument is that people should not be lab rats for the ``Chinese vaccine" (A30), manifestations against either the mobility restrictions or mandatory vaccination (A31), and insults to those who want to impose the mandatory vaccination (A32).  Topic 3 presents the largest number of tweets and the highest density (tweets/user), showing a coordinated attempt to spread specific criticisms about the ``Chinese vaccine". Topic 1 is related to the largest number of users, revealing that their central concern is mandatory vaccination. 

%% %%
\noindent \textbf{Anti-sinovaxxers.} The topics in Table \ref{tab:atributosTopicos}, together with their representative arguments in Table \ref{tab:clusterArgumentosAntisino}, confirm that the focus of this group is not only on the vaccination itself, but also the rivalry between Bolsonaro and Doria. The background behind this group's discussions is the political dispute for the 2022 Presidential election. 
As a prospective candidate, Doria leveraged the wealth of his state to support a joint effort between Instituto Butantan and the Chinese Pharmaceutical Sinovac to produce Coronavac and promoted this state program with a discourse that blends science and political interests.  In response, Bolsonaro and his supporters took a stand against the Chinese origin of Coronavac, summoning xenophobic feelings and questioning Doria's real intentions.
\textit{Topic 0} raises suspicions about the production of Coronavac and the actual intentions of Governor Doria. The largest cluster of this topic (S00) presents arguments that raise corruption suspicions connecting Coronavac and Governor Doria. The cluster S01 calls for protests against the ``Chinese vaccine", and the cluster S02 praises President Bolsonaro. Since BERTopic deploys a density-based clustering algorithm, we can see in Figure \ref{fig:matrizSimilaridadeS0} a blurred frontier between the arguments of clusters S00 and S01, while is S02 significantly distinct. 
\textit{Topic 1} highlights possible side effects of the ``Chinese vaccine" 
to support and reaffirm his rejection to Coronavac. However, we identified an interesting situation: while the arguments of the two biggest clusters in this topic denigrates the ``Chinese origins" of Coronavac, the representative argument of the third-largest cluster (S12) refutes the ones used by anti-sinovaxxers.
\textit{Topic 2} challenges the mandatory COVID-19 immunization in different states, arguing about the speed and safety of Coronavac's development. The largest cluster of arguments (S20) refutes the proposition of mandatory state-wide vaccination in São Paulo. The representative argument in cluster S21 is that Brazilians cannot be ``human guinea pigs" for a vaccine. In cluster S22, some doubts are thrown on the timely development of Coronavac regarding the allegedly brief testing period.
Topic 2 presents the highest number of tweets, associated users, and density, showing that the group's main concern is to challenge the obligation of a vaccine, based on a mixture of distrust of it origins and the real intentions of João Doria. 

\noindent \textbf{Pro-vaxxers.} According to the topics in Table \ref{tab:atributosTopicos}, users in this group are anxious to get immunized, criticize the federal government (and the President), and celebrate the results from Brazilian research centers in the development of a COVID vaccine and its approval by regulatory agencies. To understand the representative arguments in Table \ref{tab:clusterArgumentosProvax}, it is necessary to explain the right-wing agenda being implemented under Bolsonaro's government. He defends a frugal Government, proposing the privatization of several federal agencies (including health-related ones) and drastically reducing universities and research budgets. He repeatedly insisted on a federal strategy for COVID-19 management centered on scientifically-refuted ``early treatments" (e.g., chloroquine) and promoted vaccine FUD (Fear, Uncertainty, and Doubt), particularly regarding Coronavac. For instance, he claims that due to insufficient testing, Coronavac could transform people into alligators\footnote{https://thewire.in/world/jair-bolsonaro-brazil-coronavirus-vaccine-anxiety}. \textit{Topic 0} concentrates comments on the initial milestones of vaccination in Brazil, congratulating research institutes Fiocruz and Butantan, which started the production of vaccines in the national territory. The two largest clusters of arguments (P00 and P01) exalt all institutions involved with vaccination, while cluster P02 shows expectations with the beginning of the immunization.
\textit{Topic 1} addresses the enthusiasm about the beginning and new stages of the vaccination campaign against COVID, and all the arguments convey joy, excitement, and hope. The arguments of the two largest clusters (P10 and P11) celebrate the availability of vaccines in Brazil, while cluster P12 shows the expectation of users to receive the vaccine, regardless of its origin.
\textit{Topic 2} exalts the importance of science and criticizes the President's denialism. While the largest cluster (P20) concentrates on people wishing to receive the vaccine as soon as possible, the other clusters heavily criticize the president. The clusters P21 and P22 expose the President's incompetence in leading the pandemic, penalizing the population with a low supply of vaccines and consequently a prolonged crisis in health and economy.  
\textit{Topic 3} presents expectations for the availability of vaccines in Brazil, as well as criticisms to the president and his supporters, who allegedly do not make efforts to make vaccination a reality for Brazilians. The representative argument of the largest cluster on this topic (P30) is happiness about the availability of vaccines in Brazil. In contrast, clusters P31 and P32 criticize the government for making decisions seen as contrary to the fight against COVID. Cluster P31 makes fun of Bolsonaro as he claims to have supported vaccination since the early beginning, and cluster P32 criticizes the President's supporters with references to genocide. 
\textit{Topic 4} celebrates the beginning of vaccination in Brazil as a means to return to the pre-pandemic routine, in addition to criticizing the president for the actions that culminated in high levels of deaths in Brazil without any political consequences. The largest cluster on the topic (P40) expresses the expectation of vaccination for the return of schools, while the other clusters claim for vaccination to organize raids against the president (P41) and draw attention to the government's political armor, regardless of the careless actions of the President (P42). Topic 4 has the highest number of tweets and density, revealing that Brazilians are eager to get vaccinated to go back to their routine. Topic 0, which involves the largest number of users, corroborates this feeling by spreading the news on the beginning of vaccination. 

\noindent \textbf{Neutrals. } The topics identified in the Neutrals group (Table \ref{tab:atributosTopicos}) reveal they are unrest with the availability of vaccination in order to return to a normal routine without confinement. Table \ref{tab:clusterArgumentosNeutrals} shows the main arguments related to these topics considering the largest clusters. It is clear they convey the same expectations regarding the vaccination, but we did not find any reference to the government/President neither in the form of endorsement or criticism. Topic 0 questions the delay in the availability of a vaccine, using sentences such as ``Vaccine wanted" and ``Where is the vaccine?" (N00). Topic 1 gathers tweets about the COVID vaccination status around the world and comparisons with the Brazilian situation. The largest cluster of arguments on this topic makes comparisons with countries with well-defined immunization plans and are executing them (N10). Topic 2 shows a concern with the careless behavior of the population regarding masks and agglomerations. The prevalent argument is criticisms to people who participate in agglomerations, spread misinformation about the vaccine, or argue against vaccines' effectiveness (N20), but with no direct reference to the President, who often promotes these behaviors. Topic 3 comments on leisure and celebration activities when the pandemic is controlled, and the most common argument makes references to parties on the weekend (e.g., TGIF) (N30). Topic 4 reveals concerns surrounding the Brazilian immunization plan, such as which states will immunize first. The most representative arguments question when a vaccination will start in Brazil and its states (N40) and praise the Brazilian Unified Public Health System (SUS), responsible for the logistics, organization, and application of vaccines. Topic 0 has the highest density, number of users, and tweets, showing this group is engaged in expressing immunization haste. 

\noindent \textbf{Discussion. } The analysis of the topics and representative arguments enabled us to confirm our two hypotheses. First, except for the Neutrals group, all other groups intertwine their anti/pro stance with regard to vaccination with strong arguments revealing a political stance pro/anti the federal government and the President. Considering the arguments, we regard both Pro-vaxxers and Neutrals as pro-vaccination: they want to get vaccinated as soon as possible, regard immunization as the only means to return to a normal routine and criticize people's careless behavior. The Pro-vaxxers' distinctive behavior is criticisms of the President and federal government's actions and the celebration and expectation involving the beginning of vaccination against COVID-19 in Brazil. Anti-vaxxers and Anti-sinovaxxers are anti-vaccination: they do not trust vaccines without extensive research into the efficacy and side effects and consider vaccination to be an individual choice. We also confirmed that there is a difference between the Anti-vaxxers and the Anti-sinovaxxers. While the Anti-vaxxers explain their reasons for being skeptical and question the mandatory vaccination, the distinctive characteristic of the Anti-sinovaxxers is the conspiratorial claims behind the interest in a relationship with a communist country such as China and the rivalry between Doria and Bolsonaro. 

\subsection{Q2: Are these groups politically polarized?}
\label{sec:Q2}

Based on the proposed polarization index, we measured the political orientation of users belonging to the four groups. We elaborate the following hypotheses: a) the anti-vaccination stance represented by Anti-vaxxers and Anti-sinovaxxers is in essence expressed by right-oriented users are; b) the Pro-vaxxers group is composed of users who in the majority are left-oriented; c) despite its pro-vaccination stance, the Neutrals group is not politically influenced. 

\begin{figure}[!t]
\centering
\vspace{-0.25cm}
\includegraphics[width=0.5\textwidth]{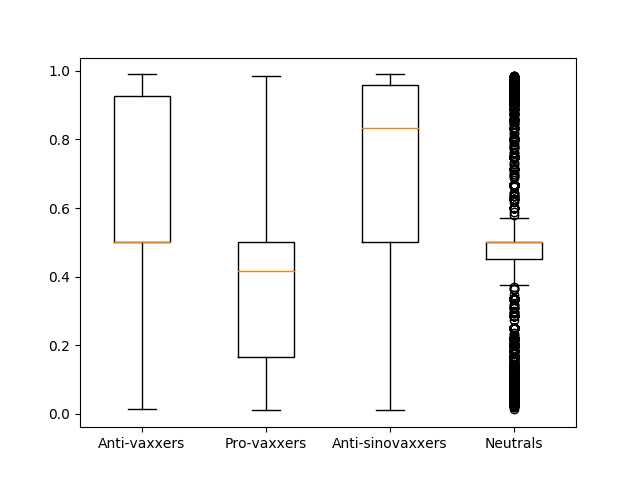}
%\caption{Polarização Política dos Usuários}
\vspace{-15pt}
\caption{Distribution of political orientation in groups}
\label{fig:boxplotPolarizacao}
\end{figure}

The boxplot in Figure \ref{fig:boxplotPolarizacao} shows the distribution of the political polarization index within each group in terms of quartiles, medians, and min/max values. In each group, we can identify a distinct political polarization trend. We can confirm that the Neutrals group is politically neutral, as the median is 50, and all users with polarization index lower/higher than 45/50 are regarded as outliers.

Despite the medians of Pro-vaxxers and Anti-vaxxers are similar, the first (Q1) and third quartiles (Q3) of these distributions reveal trends towards the left and right, respectively. 
While 75\% of Pro-vaxxers (Q3) display a polarization index of 50\% or lower, this range corresponds to only 50\% of the Anti-vaxxers. The Q1 of Pro-vaxxers is 16.66. %, and there is almost no difference between the median and Q3 values.
%Figure \ref{fig:polarizacaobarraProvax} shows the distribution of political polarization index of Pro-vaxxers, with average polarization equals to 37.46, corroborating their tendency towards the left.

Anti-vaxxers display a mirrored behavior compared to Pro-vaxxers, with 25\% of users concentrated in the range between 92.59 (Q3) and 98.97 (Max). The average index is 62.41, corroborating the tendency towards the right. Anti-sinovaxxers show a stronger polarization, given that the median is 82.27, Q3 is 95.83, and Max is 98.98. The average polarization in this group is 71.51\%. %Figures \ref{fig:polarizacaobarraAntivax} and \ref{fig:polarizacaobarraAntisino} help us to observe the trend to the right of Anti-vaxxers and Anti-sinovaxxers, respectively, from the distribution of the polarization index of the groups.

\noindent \textbf{Discussion.} This analysis enabled us to confirm all our hypotheses. It quantitatively validated the evidence of political orientation found in the previous analysis of the topics used to express their views. Although both are right-oriented, the Anti-sinovaxxers group is more extreme compared to Anti-vaxxers. We were also able to confirm that Pro-vaxxers are oriented towards the left, while the Neutrals are indeed politically neutrals. Figure \ref{fig:boxplotPolarizacao} help us to observe the trend towards the right/left of these groups.

\begin{comment}
\begin{figure}[!t]
\centering
	\includegraphics[width=0.35\textwidth]{Images/PolarizacaoProvax.png}
\caption{Political polarization of Pro-vaxxers}
\vspace{-5pt}
\label{fig:polarizacaobarraProvax}
\end{figure}

\begin{figure}[!t]
\centering
	\includegraphics[width=0.35\textwidth]{Images/PolarizacaoAntivax.png}
\caption{Political polarization of Anti-vaxxers}
\vspace{-5pt}
\label{fig:polarizacaobarraAntivax}
\end{figure}

\begin{figure}[!t]
\centering
	\includegraphics[width=0.35\textwidth]{Images/PolarizacaoAntisino.png}
\caption{Political polarization of Anti-sinovaxxers}
\vspace{-5pt}
\label{fig:polarizacaobarraAntisino}
\end{figure}
\end{comment}

\subsection{Q3: Does the polarization affect the social network structure of these groups?}
\label{sec:Q3}

The previous analyses confirmed that all three groups are politically polarized. In this section, we aim to understand if this polarization reflects on their social network structure. Our hypotheses are: a) the more polarized a group is, the more its topological structure reveals stronger connections and shorter paths for the information flow, and b) the key influencers of a network have a political alignment with the respective users.

\noindent \textbf{Graphs and Communities. }The left-hand side of Table \ref{tab:grafosecomunidades} shows the topological metrics for the group graphs. The clustering coefficients indicate that all groups have a reasonable probability of containing communities. Despite the highest clustering coefficient, the Pro-vaxxers graph has the highest number of communities, which can be explained by the scarce connections between its nodes (i.e., the smallest average degree - 5.74). The Anti-sinovaxxers is the smallest and most densely connected graph (average degree 8.31), yielding a small quantity of equally densely connected communities. In comparison, the Anti-vaxxers graph has a slightly larger number of communities and a smaller average degree, but which characterizes it as a more cohesive group compared to the Pro-vaxxers.
Recall that the Anti-vaxxers and Pro-vaxxers graphs correspond to a sample of the original groups (78,5\% and 11,8\%, respectively). 

Then, we examined each group’s communities,  seeking those with more politicians from our list. We observed that all three groups have two communities that concentrate a large number of politicians (between 120 and 154): one that concentrates on right-wing politicians and the other, left-wing politicians. All the other communities do not involve politicians or include a negligible number. These communities show that polarized users form networks with other polarized users and that these three groups have politically active segregated communities. \\

\noindent \textbf{Polarized Communities. }The right-hand side of Table \ref{tab:grafosecomunidades} highlights the properties of each group's most right/left-polarized communities. The polarized communities of the Anti-vaxxers have a comparable number of politicians. However, the right-oriented community is smaller than the left-oriented one (5 percentage points - pp), significantly more connected (average degree is 3,6 times larger), with a smaller average shortest path. This indicates that the right-oriented community has stronger connections. For the Anti-sinovaxxers, the difference of politicians in the two polarized communities is slightly bigger (16), but the metrics follow the same pattern: the right-one polarized community is smaller (12pp), with more edges (10pp), average degree 3 times larger and smaller shortest average path. The two Pro-vaxxer communities are slightly different: compared to the right-oriented one, the left-oriented community has more nodes (12pp) that are equally strongly connected (average degree is 1.7 larger). It is the polarized community with the smallest difference of politicians between both sides. 

To understand the existence of two polarized communities within each group, despite their clear right/left orientation, we examined their relationship with the politicians followed. In the right-oriented Anti-vaxxers and Anti-sinovaxxers communities, a  portion of the right-wing politicians are connected to each other (44\% and 26\%, respectively). In the left-oriented communities, there is no connection between the left-wing politicians. In other words, these left-wing politicians are followed by anti-vaccination users, while many of the right-wing politicians are not only followed but are also members of these communities/groups. In both groups, there is about 2\% of users in the right-oriented communities who are directly connected to right-wing politicians, with an average degree that is significantly superior compared to the respective community's average degree  (34.9 and 53.9 for the Anti-vaxxers/Anti-sinovaxxers, respectively). In these same communities, there is no connection of users with left-wing politicians.
In the left-oriented Anti-vaxxers and Anti-sinovaxxers communities, the politicians are not connected at all (right or left). The fraction of users connected to politicians is also very small, ranging from 0 to 0.4\%. Thus, we conclude that the left-oriented communities in these anti-vaccination groups are composed of users motivated to refute ideas from other groups, including left-oriented politicians. This finding is compatible with the arguments found in S12 and A20, which actually refute the anti-vaccination stance.

The Pro-vaxxers communities display similar trends but with different intensities, as not all right-wing partisans are against vaccination. Considering the left-oriented community, the politicians of both sides are connected to each other in a significant proportion (90\% and 29\% for left/right-wing, respectively). There is 0.5\% of users connected to right-wing politicians and 2.2\% of users connected with left-oriented ones (average degrees 3.8 and 10.7, respectively). Thus, the direct connection with left-oriented politicians is slightly superior compared to the community's average. In the right-oriented community, a significant portion of the right-wing politicians are connected to each other (64\%), and 2.4\% of users are connected to them, with an average degree of 9.8. The connections with left-wing politicians are negligible. There are two possible explanations for the presence of right-wing users and politicians in the pro-vaccine debates: confidence in science against ideologies\footnote{https://www.cnnbrasil.com.br/nacional/2021/01/23/datafolha-79-dos-brasileiros-querem-se-vacinar-contra-o-coronavirus}, or a shift in Bolsonaro's agenda to be seen as the one responsible for vaccination despite his stance in the past\footnote{https://time.com/5946401/brazil-covid-19-vaccines-bolsonaro/}, for electoral purposes.

Finally, we analyzed the main influencers of these communities, represented by the nodes with the highest in-degree and centrality values. In general, politicians are very influential in the right-oriented communities and mainstream media and science activists in the left-oriented ones. The position of the political influencers in the iGPS is displayed in Figure \ref{fig:igsp}. The users with the highest number of followers (in-degree) in right-oriented communities are politicians (Abraham Weintraub - former Ministry of Education, and General Heleno - Chief Minister of Institutional Security Office) and a right-wing journalist (Alexandre Garcia). For the left-polarized communities, in all groups, the streaming platform Netflix has the highest in-degree.

Considering the highest betweenness centrality, responsible for spreading information throughout the network, we observed for the right-oriented polarized communities two active Bolsonaro's supporters (ananiasfernanda and NiltonGNeto) and the Minister of Communications (Fabio Faria, responsible for official government communications). For the left-oriented communities, we identified an anti-Bolsonaro activist (do\_genocida), a science YouTuber (thabataganga), and a politician (GuilhermeBoulos). Considering the closeness centrality, responsible for shortening the paths to disseminate information, the right-oriented communities are related to right-wing figures: a profile that spread apps currently blocked (redpillados) and two far-right journalists (allanldsantos and RafaelFontana). For the left-oriented communities, the three figures are press-related: a media freelancer (mfox\_us), a journalist specialized in the medical area (\_FabioReis), and a news portal (g1).

\noindent \textbf{Discussion.}
Our analysis confirmed the hypothesis that the political polarization affects the social network structure. The anti-vaccination groups are composed of highly connected users, with a significant proportion of users engaged in polarized subcommunities (30\% and 34\% of the Anti-vaxxers and Anti-sinovaxxers, respectively). While the right-oriented polarized communities act as a closed bubble to reflect the echo-chamber, the left-oriented ones are composed of users pre-disposed to pierce the bubble to bounce ideas. The Pro-vaxxers display a mirrored pattern, although their network structure is more heterogeneous and open. The left-oriented polarized community engages connected politicians of both sides in different proportions, but users connected to them do not have stronger ties compared to the average users of the community. Regarding the influencers, politicians do play an active role in the anti-vaccination communities, but we also identified far-right free lancer journalists. In the Pro-vaxxers communities, in addition to politicians, we identified science popularization figures. In regards to our second hypothesis, users are connected to influencers with similar political orientation in general. However, the group is also open to refute ideas that in part are spread by influencers from the opposite political spectrum.

\begin{table*}[!t]
\caption{Properties of Groups and their Polarized Communities}
\label{tab:grafosecomunidades}
\scalebox{0.57}{
\begin{tabular}{|c|c|c|c|c|c|c|c|c|c|c|c|c|c|c|c|}
\hline
\multirow{2}{*}{\textbf{Groups}} & \multicolumn{5}{c|}{\textbf{Graph Properties}}                                                                                                & \multicolumn{10}{c|}{\textbf{Properties of the Most Polarized Community}}                                                                                                     \\ \cline{2-16} 
                                     & \textbf{\#Nodes} & \textbf{\#Edges} & \textbf{\begin{tabular}[c]{@{}c@{}}Average  \\ Degree\end{tabular}} & \textbf{\begin{tabular}[c]{@{}c@{}} Clust. \\  Coef.\end{tabular}} & \textbf{\#Com.} & \textbf{\#Nodes}                                              & \textbf{\#Edges}                                            & \textbf{\begin{tabular}[c]{@{}c@{}}Avg. Shortest   \\ Path\end{tabular}} & \textbf{Diam.} & \textbf{\begin{tabular}[c]{@{}c@{}}Average   \\ Degree\end{tabular}} & \textbf{\begin{tabular}[c]{@{}c@{}}Right-Wing  \\ Politicians\end{tabular}} & \textbf{\begin{tabular}[c]{@{}c@{}}Left-Wing  \\ Politicians\end{tabular}} & \textbf{\begin{tabular}[c]{@{}c@{}}Highest   \\ In-Degree\end{tabular}} & \textbf{\begin{tabular}[c]{@{}c@{}}Highest   Betw. \\ Centrality\end{tabular}} & \textbf{\begin{tabular}[c]{@{}c@{}}Highest   Closen. \\ Centrality\end{tabular}} \\ \hline
\textbf{{\begin{tabular}[c]{@{}c@{}}Anti-  \\ vaxxers\end{tabular}} }                & 1.558.171        & 5.252.810          & 6,74              & 0,004                                                            & 42                                                           & \begin{tabular}[c]{@{}c@{}}229.022  \\  (14,7\%)\end{tabular} & \begin{tabular}[c]{@{}c@{}}2.056.847   \\ (39,1\%)\end{tabular} & 3.56                                                                          & 9             & 17.96                                                             & 150                                                                    & 1                                                                      & AbrahamWeint                                                         & ananiasfernanda                                                                  & redpillados                                                                    \\ \cline{7-16}
& & & & & & \begin{tabular}[c]{@{}c@{}}306.478  \\  (19,7\%)\end{tabular} & \begin{tabular}[c]{@{}c@{}}859.406   \\ (16,4\%)\end{tabular} & 4.82                                                                          & 13             & 5.60                                                             & 8                                                                    & 142                                                                      & NetflixBrasil                                                         & do\_genocida                                                                  & mfox\_us                                                                    \\ \hline
\textbf{{\begin{tabular}[c]{@{}c@{}}Anti-sino  \\ vaxxers\end{tabular}} }                & 1.039.047        & 4.320.372          & 8,31              & 0,007                                                            & 38                                                           & \begin{tabular}[c]{@{}c@{}}146.407   \\ (14,1\%)\end{tabular}   & \begin{tabular}[c]{@{}c@{}}1.004.232   \\ (23,2\%)\end{tabular}  & 4.33                                                                          & 10             & 13.71                                                             & 154                                                                    & 4                                                                      & \begin{tabular}[c]{@{}c@{}}gen\_heleno\end{tabular}           & NiltonGNeto                                                                   & allanldsantos                                                                 \\ \cline{7-16}
& & & & & & \begin{tabular}[c]{@{}c@{}}270.009   \\ (26\%)\end{tabular}   & \begin{tabular}[c]{@{}c@{}}605.235   \\ (14\%)\end{tabular}  & 5.31                                                                          & 11             & 4.48                                                             & 2                                                                    & 138                                                                      & \begin{tabular}[c]{@{}c@{}}NetflixBrasil\end{tabular}           & thabataganga                                                                   & \_FabioReis                                                                 \\ \hline
\textbf{{\begin{tabular}[c]{@{}c@{}}Pro-  \\ vaxxers\end{tabular}} }                     & 2.027.338        & 5.803.832          & 5,74              & 0,003                                                            & 63                                                           & \begin{tabular}[c]{@{}c@{}}108.544   \\ (5,3\%)\end{tabular} & \begin{tabular}[c]{@{}c@{}}270.546  \\ (4,7\%)\end{tabular}  & 4.57                                                                          & 11             & 4.98                                                             & 123                                                                    & 1                                                                      & alexandregarcia                                                                    & fabiofaria                                                                       & RafaelFontana                                                                             \\ \cline{7-16}
& & & & & & \begin{tabular}[c]{@{}c@{}}356.366   \\ (17,6\%)\end{tabular} & \begin{tabular}[c]{@{}c@{}}1.510.193  \\ (26\%)\end{tabular}  & 4.68                                                                          & 16             & 8.47                                                             & 24                                                                    & 120                                                                      & NetflixBrasil                                                                    & GuilhermeBoulos                                                                       & g1                                                                             \\ \hline
\end{tabular}
}
\end{table*}

\subsection{Q4: Is there a difference in the information sources used by each group?}

\begin{figure}[!t]
	\centering
	\includegraphics[width=0.80\columnwidth]{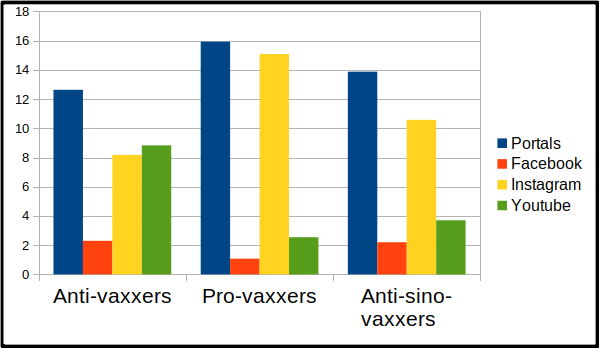}
	\vspace{-5pt}
	\caption{Use of Links per group
	\vspace{-7pt}
	\label{fig:flowLinks}}
\end{figure}

\begin{figure}[!t]
	\centering
	\includegraphics[width=0.80\columnwidth]{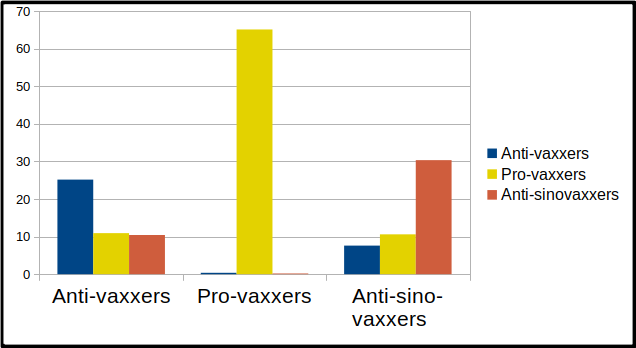}
	\vspace{-5pt}
	\caption{Mentions between groups
	\vspace{-7pt}
	\label{fig:flowMencoes}}
\end{figure}

\begin{figure}[!t]
	\centering
	\includegraphics[width=0.80\columnwidth]{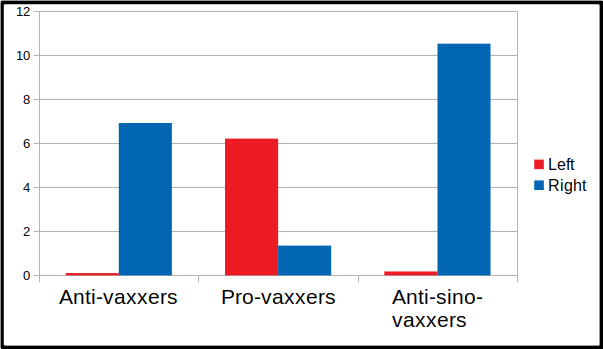}
	\vspace{-5pt}
	\caption{Mentions of politicians' tweets
	\vspace{-10pt}
	\label{fig:flowMencoesPoliticos}}
\end{figure}

% intro
% hypothesis

%In this section, we aim to understand the use of the information sources used by each group. Our hypotheses are: a) the use of links differs per group, having new portals with a higher preference in pro-vaccination stance and social media links in anti-vaccination stance; b) mentions to users of the own group are high expressing the ``echo-chamber" effect; and c) politicians have more mentions in anti-vaccination groups than in pro-vaccination one.
In this section, we aim to understand the use of the information sources by each group. Our hypotheses are: a) the anti/pro vaccination groups leverage different sources of information; b) a higher proportion of mentions to users of one's own group characterizes an ``echo-chamber" effect; and c) politicians are more influencial in anti-vaccination groups, and thus these groups tend to have a higher number of mentions to them. 

%In comparison, the use of Youtube is significant smaller among the Pro-vaxxers, who favor Instagram.

Figure \ref{fig:flowLinks} shows the proportional distribution of the web addresses collected from the tweets with URLs, according to the categories considered, i.e., news portals and the three prevalent social media platforms. News portals are the main information source in all groups, although in different proportions: 15.93\% for Pro-vaxxers, 12.63\% for the Anti-vaxxers, and 13.87\% for the Anti-sinovaxxers. The sum of social medias has a highers percentage of usage in all groups compared to news portals: 18.70\% for Pro-vaxxers, 19.31\% for Anti-vaxxers and 16.48\% for Anti-sinovaxxers. Facebook is the least used platform in all three groups, YouTube is the most popular platform among the Anti-vaxxers, and Instagram is the prevalent platform in the Pro-vaxxers and Anti-sinovaxxers groups. 

The proportional mentions per group are depicted in Figure \ref{fig:flowMencoes}, considering the tweets with mentions of each group. The Pro-vaxxers is the group that has the highest number of mentions to users of its group (65.12\%), followed by Anti-sinovaxxers (30.35\%) and Anti-vaxxers (25.16\%). Anti-vaxxers and Anti-sinovaxxers have a mirrored behavior, where Anti-vaxxers make mentions to Anti-sinovaxxers (10.42\%) and vice-versa (7.59\%). This result is expected since both represent an anti-vaccination stance and express common concerns about mandatory vaccination. In that sense, mentions to anti-vaccination users amount to 35,58\% among the Anti-vaxxers and 37,94\% among the Anti-sinovaxxers. Notice the proportion of Pro-vaxxers' mentions to Anti-vaxxers and Anti-sinovaxxers is very small (0,36\% and 0,17\%, respectively). However, these figures should be considered with caution, as the number of users in the Pro-vaxxers group is 6.4 times bigger compared to the Anti-vaxxers. 

Finally, Figure \ref{fig:flowMencoesPoliticos} depicts the proportional distribution of tweets from politicians that are used by users as sources of information and propagated. It is observed that Anti-sinovaxxers and Anti-vaxxers users propagate tweets with mentions to right-wing politicians (10.53\% and 6.93\%, respectively), while Pro-vaxxers use tweets from left-wing politicians in a smaller proportion (6,21\%). Pro-vaxxers are the ones that most spread information originated from politicians of the opposite spectrum (1,35\% for right-wing, while Anti-sinovaxxers mention 0,17\% and Anti-vaxxers 0,10\% for left-wing).

\noindent \textbf{Discussion.} 
%a) the use of links differs per group, having new portals with a higher preference in pro-vaccination stance and social media links in anti-vaccination stance
% The inspection of a sample reveal Pro-vaxxers tend to share content from verified accounts and organizations. 
Our first hypothesis was not totally confirmed. We observed a relatively similar usage of news portals among all groups, which is evidence that they all have some level of concern with the veracity of the content. Nevertheless, Pro-vaxxers have the highest usage of this type of information source (2pp and almost 4pp higher than the Anti-sinovaxxers and Anti-vaxxers, respectively). The overall usage of social media is also similar, but the groups propagate information in different ways. %The higher the proportion of the use of news portals, the higher is the concern in spreading real facts from reliable sources. 
The Anti-vaxxers have the highest usage of YouTube, a platform in which all sorts of content are freely propagated. For instance, the far-right journalists who appear with high centrality nodes in Table \ref{tab:grafosecomunidades} have their own channel on Youtube, and do not work in the mainstream media. 

%The similar usage of news portals is evidence that all groups have some level of concern with the veracity of content, but they propagate information from social networks (not always verified) in different ways. The higher the proportion of the use of news portals, the higher is the concern in spreading real facts from reliable sources. Only Anti-vaxxers have significantly higher use of social media to spread information, where YouTube, a platform in which very often conspiracy theories are freely propagated, is prevalent.

% b) mentions to users of the own group are high expressing the ``echo-chamber" effect
% It is noteworthy the mention of other groups in Anti-sinovaxxers and Anti-vaxxers, even though the users of the Pro-vaxxers group are in greater numbers, they have a percentage similar to that of the opposite anti-vaccine group. Thus, it is fair to say that the ``echo chamber" effect is observed for both stances.
% These findings are evidence that Pro-vaxxers have a high ``echo chamber" behavior, as it massively propagates information from its own group, even though it is the group that most shares information from reliable sources. As Anti-vaxxers and Anti-sinovaxxers share the same view of confrontation and political polarization, they can configure ``echo chamber" behavior in their union.

Regarding the second hypothesis, all groups display an ``echo chamber" behavior, with prevalent mentions to users of the same group. This reveals they seek to re-enforce convictions among their peers and are not very open to distinct views. Pro-vaxxers are the ones with the highest proportion of mentions to users of their own group, but they also propagate the most content from portal news. 

Finally, we confirmed the hypothesis that the anti-vaccination groups are the ones that most rely on information propagated by (right-wing) politicians, where the Anti-sinovaxxers have almost double of the proportional mentions compared to the Anti-vaxxers. Recall that proportional references to Youtube videos are higher when compared to Pro-vaxxers, reinforcing the hypothesis that they are highly politically polarized. Among the group, the Pro-vaxxers are the ones that make most mentions to politicians from a different political orientation (6.21\%, compared to 0.1 and 0.18\% of the anti-vaccination groups).

% c) politicians have more mentions in anti-vaccination groups than in pro-vaccination one.

%Anti-sinovaxxers and Anti-vaxxers have a greater interest in spreading information of politicians oriented to their political polarization index, while Pro-vaxxers look to other sources and try not to close in on the followers of right-wing politicians. This is evidence that Anti-sinovaxxers are the most politically polarized and most influenced by information with a political bias, followed by the Anti-vaxxers. Tweets using left-wing politicians are considerably less. As for the Pro-vaxxers, although the percentage of mentions of left-wing politicians is higher, the difference is not so great. 

\section{Threats to Validity}\label{sec:threats}
This section discusses the threats to the validity of our study. One of the main threats is the way the groups were defined. The use of hashtags \footnote{https://help.twitter.com/en/using-twitter/how-to-use-hashtags} for automatic collection of groups on social networks is a widely used form for this purpose, however, it can lead to different types of bias. First, the hashtags may not represent the target population, and we mitigated this risk by a  careful inspection of frequent hashtags. Another risk is that tweets might be falsely inserted in the context of a hashtag because they refute an idea represented by the hashtag (false positives). The detailed investigation of topics revealed this bias was inserted in our users, as revealed by clusters S12 and S20. However, the amount of users is small within the total of users that compose each group and should not affect the overall patterns identified. 

Another threat of validity is the selected politicians to represent the political polarization. We mitigate the risk by using iGPS, which is based on a widely consolidated statistical model \cite{Barbera2015}. Another threat is users who are politically active but follow politicians of both sides, and thus are perceived as neutral. More complex models can be evaluated in the future, which consider, in addition to the followed politicians, the textual contents of the tweets as a refinement \cite{PreotiucPietro2017}. 

Finally, it is common knowledge that the Twitter audience may not represent characteristics of the general population, especially in analyzes such as this, which represents a frame of the public of this social network.

\section{Conclusions and Future Work} 
\label{sec:conclusoes}

This paper investigated the influence of political polarization on the vaccination stances expressed by Brazilians on Twitter. We used a multi-dimensional framework that encompasses concerns, measurement of polarization, social network properties, and information sources. We identified three polarized groups. Pro/against stances were politically polarized towards the left/right, respectively, where the anti-sinovaxxers were the most politically polarized.  

%In this paper, we proposed a framework to analyze how political polarization affects the behavior of groups with opposed stances, using the Brazilian COVID polarized scenario as case study. Our framework leverage techniques to characterize group behavior in terms of multiple dimensions: demographics, concerns, psychological aspects, social network properties and information sources. We used these properties to explain the  behavior of Chloroquiners and Quarenteners in terms of the Identity Protective Cognitive theory. The same analysis framework can be used to investigate the implications of political polarization on other controversial topics that are frequently affected by political polarization, such as environment protection, vaccine hesitancy, among others.

Our results contradict studies that did not observe a political bias in anti-vaccination behavior \cite{Hornsey2018,Czarnek2020}, but our analysis is restricted to the specific Brazilian COVID scenario. The anti-vaccination groups in Brazil express concerns beyond the observed hesitation highlighted in international studies (e.g., \cite{Burki2020,Cinelli2020}). 
Our analysis provided evidence that the opposing groups have political motivations, given that two candidates for the 2022 presidential election are using COVID immunization as an electoral platform. 

The Anti-sinovaxxers is the most polarized group.%, and its users demographics are aligned with Bolsonaro's voters. 
Their concerns relate to the vaccine's origin and conspiratorial issues regarding Doria's political intentions. Anti-vaxxers and pro-vaxxers have opposed political polarization: their indexes orbit in opposite directions the neutrality with a similar distance, and the expressed concerns diverge on the importance of collective immunization. All groups have two polarized communities, segregated in terms of right/left politicians followed. In general, while one acts as a closed bubble to reflect the echo chamber, the other is composed of users pre-disposed to pierce the bubble to bounce ideas. The right-oriented polarized communities of the anti-vaccination groups are more densely connected, and in general, are more influenced by politicians. Pro-vaxxers and Neutrals share haste for immunization, perceived as the only means to get back to a normal routine, but pro-vaxxers criticize the government's actions. An ``echo chamber" effect was observed in all groups, mostly propagating ideas aligned with their own point of view.

As future work, we want to expand the range of dimensions analyzed by the framework, such as historical behavior analysis.% and new techniques for identifying political polarization. 
We also want to expand the concepts of the dimensions already present, such as information flow in the groups and exploring new metrics to characterize the social network communities, and  understand the dynamics of each interest group.

%References and End of Paper
%These lines must be placed at the end of your paper
\bibliographystyle{aaai}
\bibliography{aaai.bib}
\end{document}